\documentclass[conference]{IEEEtran}
\IEEEoverridecommandlockouts
\usepackage{cite}
\usepackage{amsmath,amssymb,amsfonts}
\usepackage{algorithmic}
\usepackage{graphicx}
\usepackage{textcomp}
\usepackage{xcolor}
\usepackage{pgf}
\usepackage{tikz}
\usetikzlibrary{arrows,automata}
\usepackage[latin1]{inputenc}
\usepackage{verbatim}
\usepackage{subcaption}

\begin{document}

\title{Control of Status Updates for Energy Harvesting Devices that Monitor Processes with Alarms}

\author{
	\IEEEauthorblockN{George Stamatakis\IEEEauthorrefmark{1},
		Nikolaos Pappas\IEEEauthorrefmark{2},
		Apostolos Traganitis\IEEEauthorrefmark{1}
		\IEEEauthorblockA{\IEEEauthorrefmark{1} Institute of Computer Science, Foundation for Research and Technology - Hellas (FORTH)}
		\IEEEauthorblockA{\IEEEauthorrefmark{2} Department of Science and Technology, Link\"{o}ping University, Campus Norrk\"{o}ping, Sweden}
		E-mails: \{gstam,tragani\}@ics.forth.gr, nikolaos.pappas@liu.se}
}

\maketitle

\begin{abstract}
In this work, we derive optimal transmission policies in an energy harvesting status update system.
The system monitors a stochastic process which can be either in a normal or in an alarm state of operation.
We capture the freshness of status updates for each state of the stochastic process by introducing two Age of Information (AoI) variables and extend the definition of AoI to account for the state changes of the stochastic process.
We formulate the problem at hand as a Markov Decision Process which, under the assumption that the demand for status updates is higher when the stochastic process is in the alarm state, utilizes a transition cost function that applies linear and non-linear penalties based on AoI and the state of the stochastic process. Finally, we evaluate numerically the derived policies and illustrate their effectiveness for reserving energy in anticipation of future alarm states.
\end{abstract}

\section{Introduction}
Providing timely status updates is of crucial importance to applications that offer monitoring services in cyber-physical systems~\cite{stankovic2014}. These applications are the cornerstone of the smart infrastructure, which is being enabled by the Internet of Things (IoT). Examples of such applications include, but are not limited to, smart cities, smart factories and grids, smart agriculture, parking and traffic management, e-Health and environment monitoring~\cite{stankovic2014}. The proliferation of these applications is expected to have a profound impact on key sectors of economy, and this has spurred research on their particular operational requirements~\cite{LMT2014}.

A key result in the field was the realization that the objective of timely status updating is not captured by metrics such as throughput and delay. To alleviate this problem the authors in~\cite{kaul2012real} introduced a new metric called Age of Information (AoI). Ever since its introduction the problem of optimally deciding on status update generation and transmission in order to minimize AoI based metrics has received much attention from the research community \cite{SunISIT2017, SunTIT2017, Stamatakis2018}. AoI has been extended to other metrics such as the value of information, cost of update delay, and non-linear AoI \cite{kosta2017age, sun2018sampling, zheng2019closed}.

Another significant problem in the field, is the choice of an appropriate energy source for the remote sensors~\cite{EHIoTCommag15}. Batteries have a limited life span and replacing them can turn out to be extremely costly as there may be hundreds of sensors in remote or even unreachable locations. To address this problem energy harvesting (EH) technologies have been developed to supply the necessary power to remote sensors~\cite{akan2017internet}. In any case, when EH and/or batteries are used, stored energy must be utilized with reservation so that there will be enough available when it is most needed.

In \cite{BacinoglouITA2015}, the problem of optimizing the process of transmitting updates from an EH source to a receiver to minimize the time average age of updates is considered. Similar studies can be found in \cite{lazy_timely, ArafaGC2017, age_eh3, ArafaICC2018, WuYangTGCN18, age_eh1}. The optimal sampling policy for IoT EH-enabled devices that minimizes the long-term weighted sum-AoI is investigated in \cite{Abd-Elmagid2018}. In \cite{Krikidis2018}, the performance of a wireless power transfer sensor network in terms of the average AoI was studied. The interplay of throughput/delay and AoI in the two-user multiple access channel with one EH-source was studied in \cite{age_ZC}. The average AoI is studied in \cite{EH_DRL_AoI2019} for status updates sent from an EH transmitter with a finite-capacity battery. The optimal scheduling policy is studied when the channel and energy harvesting statistics are known. For the case of unknown environments, the authors proposed an average-cost reinforcement learning algorithm that learns the system parameters and the status update policy in real time.

In this work, we show that the problem of reserving energy for use when it is most needed can be closely related to the problem of timely status updates. 
More specifically, we consider an EH status update system that monitors a stochastic process which can be in two states.
The first state corresponds to a \emph{normal} state of operation while the second one corresponds to an \emph{alarm} state of operation.
Under this paradigm fall many systems where, under normal operation, events occur with some probability at specific time intervals, while under alarm operation, events occur with a much higher probability.
As an example consider the rate of packet arrivals in a network under normal operation and the rate of packet arrivals under a period of denial of service attack.
Furthermore, in this work we focus on systems where the demand for fresh status updates during alarm operation is much higher compared to that during normal periods of operation.
To cope with the increased demand for status updates during such periods the system has to consider the characteristics of the energy arrival process as well as to reserve energy when possible. 

To the best of our knowledge this is the first work to consider an AoI based status update system for a two-state stochastic process and study the impact of constrained energy resources on the optimal status update transmission policies.
Furthermore, to efficiently model the problem at hand we introduce two AoI variables, one for each state of the stochastic process and extend the definition of AoI in order to include cases where the state of the stochastic process has changed while the monitoring application has not been informed of that change.
Finally, we present results that indicate the impact on the optimal policy of the probability to harvest energy, the probability to successfully transmit status updates, and the probability for the monitored process to change state given its current state.

\section{System Model}
\label{sec:systemModel}
\begin{figure}[h]
	\centering
	\includegraphics[scale=0.7]{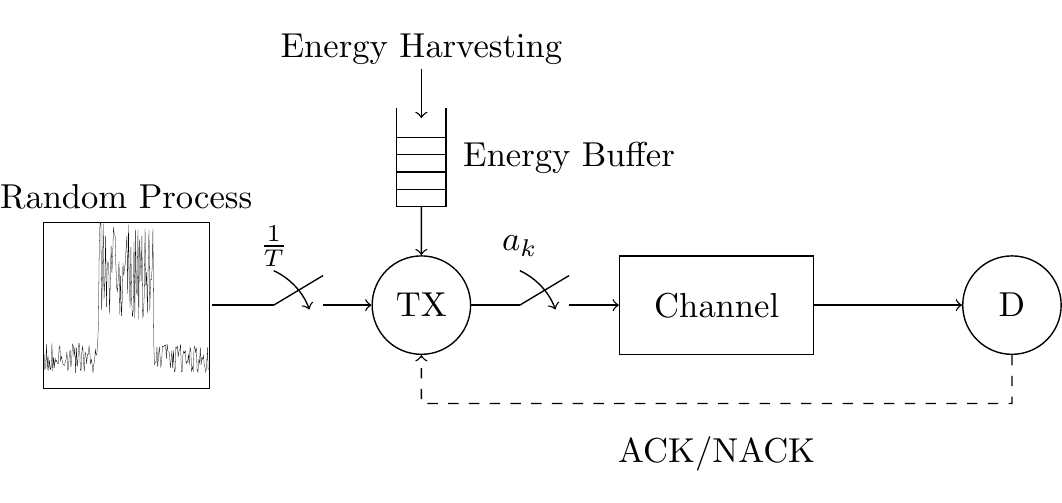}
	\caption{An EH status update system for a stochastic process with normal and alarm states of operation.}
	\label{fig:systemDiagram}	
\end{figure}

\begin{figure}[h!]
	\centering
	\includegraphics[scale=0.75]{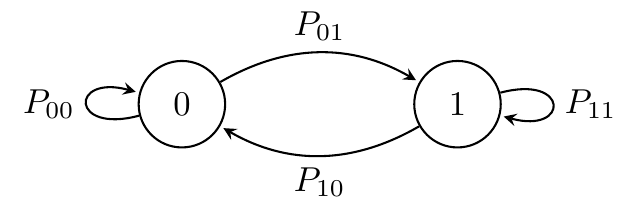}
	\caption{The states of the stochastic process and transition probabilities.}
	\label{fig:monitoredSystem}
\end{figure}

The system we consider is presented in Fig.~\ref{fig:systemDiagram} and consists of an energy harvesting (EH) sensor that monitors a stochastic process and transmits status updates at a destination node $D$.
We assume that $D$ is one-hop away from the sensor, that time is slotted, and each time-slot has duration $T$.
At the beginning of each time-slot the stochastic process can be in one of two states. 
The first state, denoted with $0$, corresponds to a \emph{normal} state of operation while the second state, denoted with $1$, corresponds to an \emph{alarm} state of operation.
An example of such a process appears in Fig.~\ref{fig:systemDiagram}.
We expect that a monitoring application for such a stochastic process should provide more frequent status updates during alarm periods of operation.
Let $\{Z_k\}$, $k=0, 1, \dots$ be the sequence of the stochastic process's states over time. Then we assume that the state of the stochastic process remains constant for the duration of a time-slot. At the beginning of the ($k+1$)-th time-slot, the state of the stochastic process will change from $Z_k = z$ to $Z_{k+1} = z'$ according to transition probabilities $P_{zz'}$, $z, z' \in \{0,1\}$, as shown in Fig.~\ref{fig:monitoredSystem}.

At the beginning of each time-slot the sensor generates a fresh status update and, subsequently, it decides whether to transmit it to the destination or not. The sensor is equipped with an energy buffer that can store an integer number of energy units up to a maximum capacity of $E_{max}$ energy units. The sensor harvests an energy unit in a time-slot with probability $P_e$.
We assume that each status update transmission consumes a single energy unit and that no transmission can occur if the buffer is empty.
In this work, we disregard energy costs related to other functions of the sensor such as sensing, processing, and storing data in memory. Each transmission may succeed with probability $P_s$, independently of the transmission outcomes of previous transmissions.
Furthermore, we assume that a packet transmission is acknowledged instantly.

We utilize the AoI metric to characterize the freshness of status updates at the destination.
AoI was defined in \cite{kaul2012real} as the time that has elapsed since the generation of the last status update that has been successfully decoded by the destination.
However, in this work, we have to also consider the state changes of the stochastic process and the fact that the destination will be unaware of any such state change until the successful reception of a fresh status update.
Furthermore, the sensor node has to effectively decide whether to transmit a fresh status update considering both the increased (decreased) demand for status updates when the stochastic process is in an alarm (normal) state and the constrained energy resources stored in the buffer. 
To achieve its goal, the sensor should utilize its knowledge of both the change in the stochastic process' state and the value of AoI at the destination.

To accommodate for this case, we make use of two separate AoI variables, one for each state of the stochastic process. 
We denote with $\Delta_k^z, z\in\{0,1\}$ the AoI that corresponds to the $z$-th state of the stochastic process at time $k$.
Furthermore, we denote the sequence of time indices where a state change has occurred as $\{\tau_n: Z_{\tau_n} \neq Z_{\tau_n - 1},\ n=1, 2, ...\}$ and define $\tau_N$ to be the time index of the most recent state change for the stochastic process by time $k$, i.e., $N = \max \{n : \tau_n < k\}$. 
Finally, let $Z_k^d$ denote the state that the destination \emph{knows} as the process's state at time $k$, i.e., the state of the stochastic process included in the last status update received by the destination.
Then we define AoI as follows, 
\begin{equation}
\label{eq:AoiDefinition}
\Delta_k^z = \begin{cases}
k - U_k, & \text{if $z = Z_k^d$}, \\
\min\{k - \tau_N, \Delta_{max}^z\}, & \text{if $z \neq Z_k^d$ and $z = Z_k$} \\
0, & \text{if $z \neq Z_k^d$ and $z \neq Z_k$} 
\end{cases}
\end{equation}
where $U_k$ is the time-stamp of the last packet that has been received at the destination by time $k$ and $\Delta_{max}^z$ is the maximum value of AoI corresponding to the maximum level of staleness.

The first branch of (\ref{eq:AoiDefinition}) applies to the AoI variable that corresponds to the state of the stochastic process that is known at the destination at time $k$ and coincides with the definition of AoI in~\cite{kaul2012real}.
The second branch of (\ref{eq:AoiDefinition}) applies in the case where \emph{one or more state changes} have occurred and, as a result, the current state of the stochastic process is different from the one known at the destination ($Z_k \neq Z_k^d$). 
In such a case the AoI for $z = Z_k$, i.e., $\Delta_k^z$, is defined to be equal to the time that has elapsed since the last state change ($\tau_N$).
Finally, the third branch of the equation applies for AoI $\Delta_k^z$ when $z$ is neither the state known by the destination nor the currently active state, i.e., the state known to the destination at the $k$-th time-slot, $Z_k^d$, is equal to the true state of the stochastic process. In such a case, the state $z$ that is not currently active, i.e., $z \neq Z_k^d$, is assigned an AoI value of zero.

\begin{figure}[htb!]
    \centering
    \begin{subfigure}[b]{\columnwidth}
    	\centering
    	\includegraphics[scale=0.5]{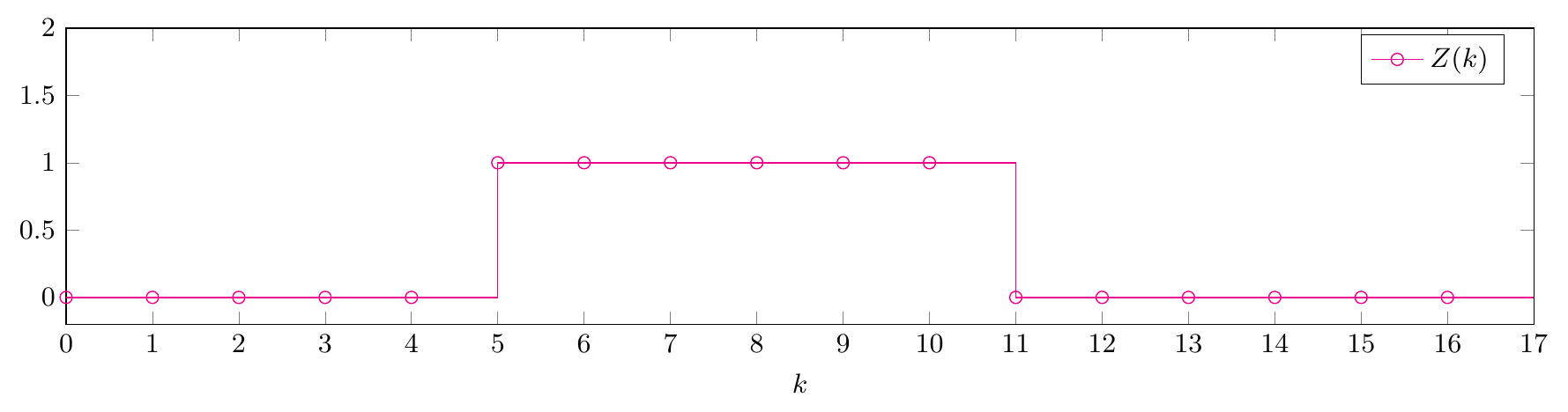}
	   	\caption{Evolution of the stochastic process' state over time.}
	   	\label{fig:randomProcessStateEvolution}	
   	\end{subfigure} \\
    \begin{subfigure}[b]{\columnwidth}
    	\centering
    	\includegraphics[scale=0.5]{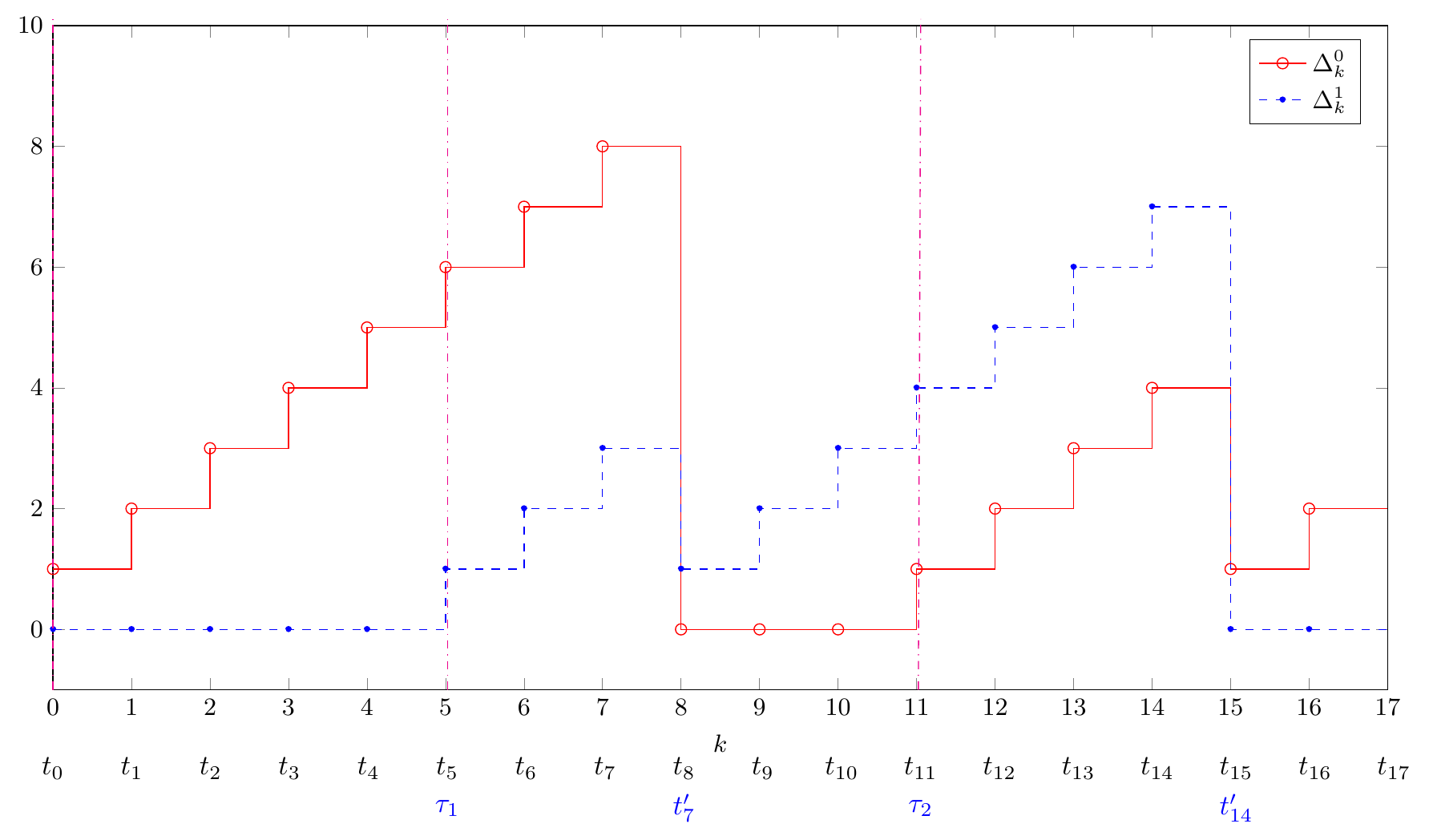}
    	\caption{Evolution of $\Delta_k^0$ and $\Delta_k^1$ over time for the $Z_k$ presented above.}
    	\label{fig:AoI_and_cost_function}
    \end{subfigure}
\caption{The first sub-figure presents the evolution of the stochastic process' state over time. The second sub-figure presents the evolution of the AoI for each state of the stochastic process.}
\label{fig:AoiDefinition}
\end{figure}

An example of the evolution of variables $\Delta_k^0$ and $\Delta_k^1$ over time appears in Fig.~\ref{fig:AoiDefinition}. We indicate the time instances where status updates are generated with $t_k,\ k \geq 0$. The time instances where the status update generated at the $k$-th time-slot successfully arrives at the destination are indicated with $t_k^{'}$. Finally, we use $\tau_c, c\geq 0$ to indicate the times where the state of the stochastic process changes. At time $k=0$, we assume that the destination receives a status update indicating that the stochastic process is in state $Z(0) = 0$ and, as a consequence, $\Delta^0$ is incremented by one at each time-slot, while $\Delta^1$ remains zero. 
At time $k=5\ (\tau_1)$ the state of the stochastic process changes to $Z(5)=1$ while the destination has not yet received a status update with the relevant information.
Thus, both  $\Delta_k^0$ and $\Delta_k^1$ are incremented after each time-slot according to the first and second branch of (\ref{eq:AoiDefinition}) respectively.
At time $k=8$ the status update generated at the previous time-slot is received by the destination which is informed that the true state of the stochastic process is $1$ and that this happened at $\tau_1=5$. 
From that point on, $\Delta_k^0$, is set to zero according to the third branch of (\ref{eq:AoiDefinition}) while $\Delta_k^1$ is incremented according to the first branch of (\ref{eq:AoiDefinition}).
Finally, at $k = 11 (\tau_2)$, the state of the stochastic process changes from $0$ to $1$ and the same process is repeated.

\section{Problem Formulation}
\label{sec:problemFormulation}
In this section we present the state, action and random variable spaces of the system as well as the system's transition and cost functions.

\subsubsection*{State Space}
At the beginning of the $k$-th time-slot the state of the system is represented by the following state vector,
\begin{equation}
s_k = [Z_k, Z_k^d, E_k, \Delta_k^0, \Delta_k^1]^T
\end{equation}
where $Z_k\in\{0, 1\}$ denotes the state of the stochastic process, $Z_k^d$ is the state of the stochastic process known by the destination at $k$, $E_k = 0, 1, ..., E_{max}$ is the energy stored in the energy buffer, $\Delta_k^z, z\in\{0,1\}$ is the AoI at the destination for the $z$-th state of the stochastic process and $T$ is the transpose operator. 
We denote the set of all system states as $S$.

\subsubsection*{Actions}
Given that there exists at least one energy unit stored in the buffer the sensing node has the choice of either transmitting a fresh status update or eschewing the transmission and reserve energy for future use.
We denote with $a_k \in \{0,1\}$ the action taken by the sensing node, where $0$ indicates the action of not transmitting a status update and $1$ indicates the action of transmitting a status update. 
If the energy buffer is empty the sensor is constrained to action $0$. 
Finally, we use $a_s^*$ to denote the optimal action in a state $s$, $A$ to denote the set of all actions and $A(s)$ to represent the set of admissible actions at state $s$.
\subsubsection*{Random variables}
Given the state of the system and the action taken by the sensor the system will make a stochastic transition to a new state. The transition to this new state is determined by the realization of three random variables.
The fist one is $W_k^s \in \{0,1\}$ and represents the random event of a successful transmission over the noisy channel. 
We assume that a transmission is successful with probability $P_s$.
If the sensor decides against a transmission at time-slot $k$ then $W_k^s$ will be zero with probability one.
The second random variable is $W_k^e \in \{0,1\}$ and represents the random event of a energy unit arrival.
We assume that the sensor will harvest an energy unit in the duration of a time-slot with probability $P_e$.
The final random variable is $W_k^z \in \{0,1\}$ which represents the new state of the random process. 
The value of $W_k^z \in \{0,1\}$ is determined by the transition probabilities presented in Fig.~\ref{fig:monitoredSystem}.
We assume that the values of all random variables become known to the sensor at the very end of the $k$-th time-slot as is typical in optimal control theory~\cite{B12}.
Finally, we assume that $W_k^s$, $W_k^e$, and $W_k^z$ are independent random variables and, furthermore, that their values are independent of the values in previous time-slots and identically distributed over all time-slots.
We use the random column vector $W_k = [W_k^s, W_k^e, W_k^z]^T$ to collectively refer to the random variables of the system.

\subsubsection*{System Dynamics}
\label{sec:systemDynamics}
The next system state, $s_{k+1}$ will be given by the values of $Z_{k+1}$, $E_{k+1}$, and $\Delta_{k+1}^z, z\in\{0,1\}$.
More specifically the state of the stochastic process at the $(k+1)$-th time-slot is provided by the random variable $W_k^z$ whose value becomes known by the end of the $k$-th time-slot. 
\begin{equation}
Z_{k+1} = W_k^z,
\end{equation}
while the state of the stochastic process known by the destination assumes a new value only in the case of a successful status update transmission,
\begin{equation}
Z_{k+1}^d = \begin{cases}
Z_k^d, & \text{if $W_k^s = 0$} \\
Z_k, & \text{if $W_k^s = 1$}.
\end{cases}
\end{equation}
The energy stored in the energy buffer at the beginning of the $(k+1)$-th time-slot depends on whether the sensor transmitted a status update and an energy unit was harvested during the $k$-th time-slot,
\begin{align}
E_{k+1} = \begin{cases} 
	E_k + W_k^e -1, & \text{if $a_k = 1$} \\
	E_k + W_k^e, & \text{if $a_k = 0$}.
	\end{cases}
\end{align}
Here, we present a recursive definition for $\Delta_{k+1}$, although the evolution of the AoI variables over time was described in (\ref{eq:AoiDefinition}),
\begin{align}
\Delta_{k+1}^z = \begin{cases}
0, & \text{if $z \neq Z_k$ and $z \neq Z_k^d$} \\
1, & \text{if $W_k^s = 1$ and $z = Z_k$} \\
\min\{\Delta_k^z + 1, \Delta_{max}^z\}, & \text{if ($z = Z_k$ and $z \neq Z_k^d$) or} \\
 				 &\quad \text{ ($W_k^s = 0$ and $z = Z_k$)}.\\
\end{cases}
\end{align}

\subsubsection*{Transition cost function}
The cost associated with each state transition is given by,
\begin{align}
\label{eq:transitionCost}
g(s_k, a_k, w_k) &= \big(1-Z_k\big)\cdot \Delta_k^0 + Z_k \cdot \big( \Delta_k^1\big)^2
\end{align}
where $w_k$ is the realization of random vector $W_k$ at the $k$-th time-slot.
From (\ref{eq:transitionCost}) we observe that when the stochastic process is in the normal state ($Z_k = 0$), the transition cost increases linearly with $\Delta_k^0$, while when the system is in the alarm state ($Z_k = 1$) the transition cost increases with the square of $\Delta_k^1$.
Thus, the transition cost function captures the increased demand for status updates when $Z_k = 1$.\footnote{Here we consider the square function to facilitate presentation. However, we could use any function that can model the case of aging faster than linearly.}

\subsubsection*{Total cost function}
We are interested in minimizing the total cost accumulated over an infinite time horizon,
\begin{equation} 
\label{eq:cummulativeCostFunction}
J_{\mu}(s_0) = \underset{N \to \infty}{\lim} \underset{\underset{k=0,1,\dots}{W_k,}}{\mathop{\mathbb{E}}} \left\lbrace \sum_{k=0}^{N-1} \gamma^k g(s_k, a_k, w_k) | s_0\right\rbrace,  
\end{equation}
where $s_0$ is the initial state of the system, expectation $\mathbb{E}\lbrace \cdot \rbrace$ is taken with respect to the joint probability distribution of random variables $W_k$, $k=0, 1, \dots$ and $\gamma$ is a discount factor, i.e., $0 < \gamma < 1$, indicating that the importance of the induced cost decreases with time.
Finally, let $\mu$ represent a deterministic policy 
that maps each state to a specific action. 

\emph{Our objective} is to find an optimal policy $\mu^*$ that minimizes (\ref{eq:cummulativeCostFunction}).

\section{Optimal Policy}
\label{sec:optimalPolicy}

The dynamic program presented in section~\ref{sec:problemFormulation} is characterized by finite state, control, and probability spaces. 
The transitions between states depend on $s_k$, $a_k$, and $w_k$ but not on their past values and, additionally, the probability distribution of the random variables is invariant over time. 
The cost associated with a state transition is bounded and the cost function $J(\cdot)$ is additive over time.
Due to these structural properties the dynamic system at hand constitutes a Markov Decision Process (MDP)~\cite{B12} whose dynamics are fully captured by its state transition probabilities,
\begin{multline}
\footnotesize
\label{eq:MdpTransitionProbabilities}
p_{ij}(u) = P \lbrace s_{k+1}=j|s_k=i,\ a_k=a \rbrace =  \\
\sum_{(w_k^s, w_k^e, w_k^z) \in W_j} \hspace{-13px} P \lbrace W_k^s = w_k^s \rbrace P \lbrace W_k^e = w_k^e \rbrace P \lbrace W_k^z = w_k^z \rbrace	 
\end{multline}
where, $s\in S,\ a\in A(s),\ (w_k^s, w_k^a, w_k^z) \in \{0,1\}^3$ and  $W_j = \{(w_k^s, w_k^e, w_k^z) \in \{0, 1\}^3: j = f(i, u, [w_k^s, w_k^e, w_k^z]^T)\}$ and $f(\cdot)$ is the system state transition function that encompasses all state variable transitions presented in Section~\ref{sec:systemDynamics}.
For the MDP under consideration, given that $0<\gamma<1$, there exists an optimal stationary policy $\mu^*$
which is independent of the initial state of the system and deterministic~\cite[Sec. 2.3]{B12}, i.e., whenever the system is in state $i$, $\mu^*(i)$ always applies the same control $u$ that minimizes (\ref{eq:cummulativeCostFunction}), i.e., 
\begin{equation}
\mu^* = \arg \underset{\mu \in \mathcal{M}}{\min} J_{\mu}(i),\qquad \text{for all $i \in S$},
\end{equation}
where $\mathcal{M}$ is the set of all policies.
Let $J^*(i)$ be the infinite horizon discounted cost attained when the optimal policy $\mu^*$ is applied and the system begins at state $i$. $J^*(i)$ satisfies the Bellman equation,
\begin{align}
\label{eq:BellmanMDP}
J^*(i) &= \underset{u\in U(i)}{\min} \sum_{j = 1}^n p_{ij}(u) \left [g(i,u, j)+\gamma J^*(j) \right ],\text{ for all $i \in S$}, 
\end{align}
where $n$ is the cardinality of the state space. 
Given that the transition cost $g(i, u, j)$ is bounded and that $0< \gamma < 1$ the operator, 
\begin{equation}
\label{eq:VI}
(TJ)(i) = \underset{u\in U(i)}{\min} \sum_{j = 1}^n p_{ij}(u) \left [g(i,u,j)+\gamma J(j) \right ],
\end{equation} 
is a contraction mapping~\cite[Assumption D, Prop. 1.2.1, pg.14]{B12} and starting with an arbitrarily initialized vector $J(i), i \in S$, and repeatedly applying transformation $(TJ)$ for all states $i \in S$ we attain the optimal cost $J^*$ and at the same time derive the optimal policy $\mu^*$ for all $i\in S$ according to~\cite[Prop. 1.2.1, pg.14]{B12} which states that,
\begin{equation}
J^*(i) = \lim_{m\rightarrow\infty} (T^mJ)(i), 
\end{equation} 
where $(T^mJ)(i) = (T(T^{m-1}\dots(T^0 J))(i)$ and $(T^0J)(i) = J(i)$.
(\ref{eq:VI}) is a formal description of the Value Iteration (VI) algorithm~\cite[Section 2.2, pg. 84]{B12}. 
In this work, we utilize VI in order to provide insight into the operation of the system and the characteristics of its optimal policy.

\section{Results}
In this section, we evaluate numerically the infinite horizon discounted cost, $J^*(\cdot)$, of the optimal policy for various configurations of the system's parameters.
Throughout all experiments we set the discount factor $\gamma$ to be equal to $0.99$, the AoI counters' upper bounds, $\Delta_{max}^0$ and $\Delta_{max}^1$ are both set equal to $10$ and, in order to facilitate intuition, we assume that the initial state of the system, $s_0$, is always the same.
More specifically, we assume that the sensor is deployed when the random process is in a normal state $(Z_k = 0)$, that this is known to the destination $(Z_k^d = 0)$ and that the energy buffer is initially empty ($E_k = 0$). 
AoI counters $\Delta_k^0$ and $\Delta_k^1$ are respectively set to values $1$ and $0$ resulting in $s_0 = [0, 0, 0, 1, 0]^T$.

\begin{figure}[!htb]
	\centering
	\includegraphics[scale=0.9]{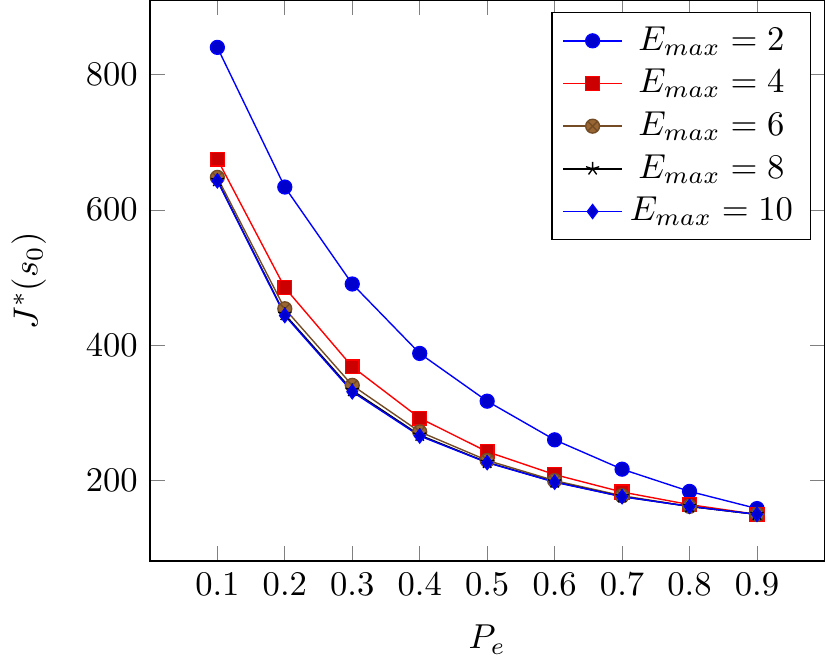}
	\caption{Impact of energy buffer's capacity, $E_{max}$ on $J^*(s_0)$.}
	\label{fig:ehProbVsCostVsBufferSize}
\end{figure}
In Fig.~\ref{fig:ehProbVsCostVsBufferSize} we present $J^*(s_0)$ for different values of the capacity of the energy buffer $E_{max}$, as the probability to harvest energy, $P_e$, varies. 
In all experiments of Fig.~\ref{fig:ehProbVsCostVsBufferSize} the stochastic process' state transition probabilities were set to
\begin{equation}
P_z = \begin{bmatrix}
\label{eq:transitionProbabilityMatrix}
0.9 & 0.1 \\
0.2 & 0.8 
\end{bmatrix}
\end{equation}
and the transmission success probability $P_s$ was set to $0.8$. 
Fig.~\ref{fig:ehProbVsCostVsBufferSize} depicts that both being in an environment where harvesting energy occurs with high probability and having an energy buffer with large capacity have a positive effect in reducing $J^*(s_0)$.
The results in Fig.~\ref{fig:ehProbVsCostVsBufferSize} also suggest that the impact of the energy buffer's capacity on $J^*(s_0)$ is negligible when $E_{max}$ grows beyond a certain bound for the given system setup.

\begin{figure}[!htb]
	\centering
	\includegraphics[scale=0.9]{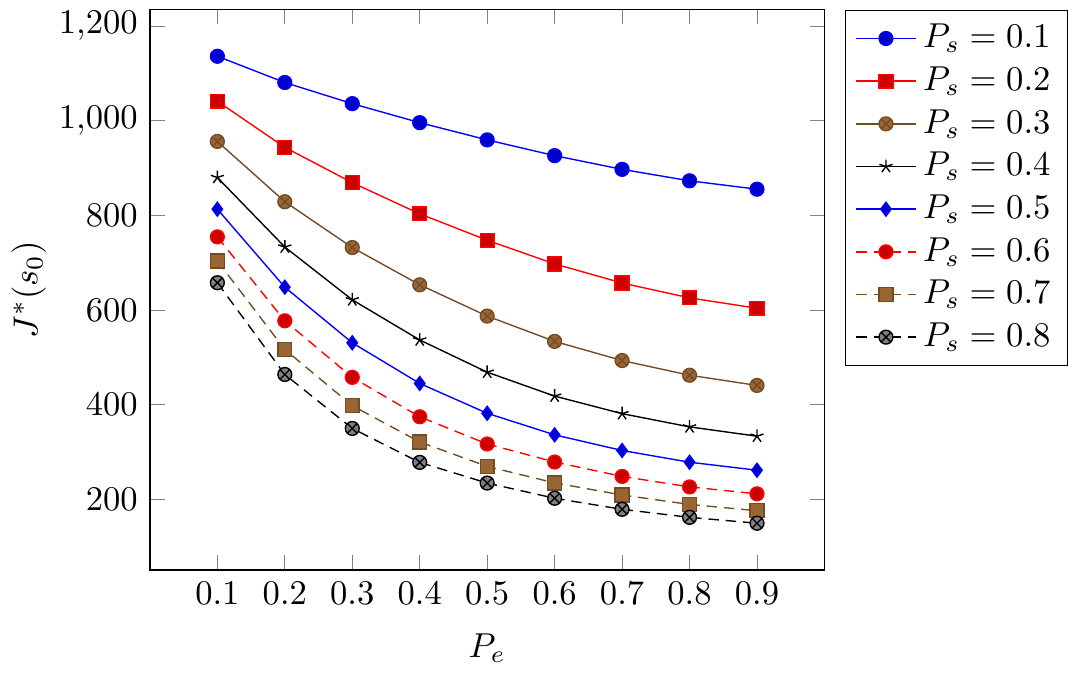}
	\caption{Impact of transmission success probability $P_s$ on $J^*(s_0)$.}
	\label{fig:ehProbVsCostVsTxSuccessProb}
\end{figure}
Fig.~\ref{fig:ehProbVsCostVsTxSuccessProb} presents $J^*(s_0)$ as a function of $P_e$, for different values of the transmission success probability $P_s$. In this set of experiments, $P_z$ is the matrix $P_z$ defined in (\ref{eq:transitionProbabilityMatrix}) and $E_{max}$ was set to $5$.
Fig.~\ref{fig:ehProbVsCostVsTxSuccessProb} depicts that increasing $P_s$ will always result in a lower $J^*(s_0)$ value. 
Furthermore, the results suggest that, given the capacity of the energy buffer, one should aim for a higher $P_s$ value in environments with low probability to harvest energy.

\begin{figure}[!htb]
	\centering
	\includegraphics[scale=0.9]{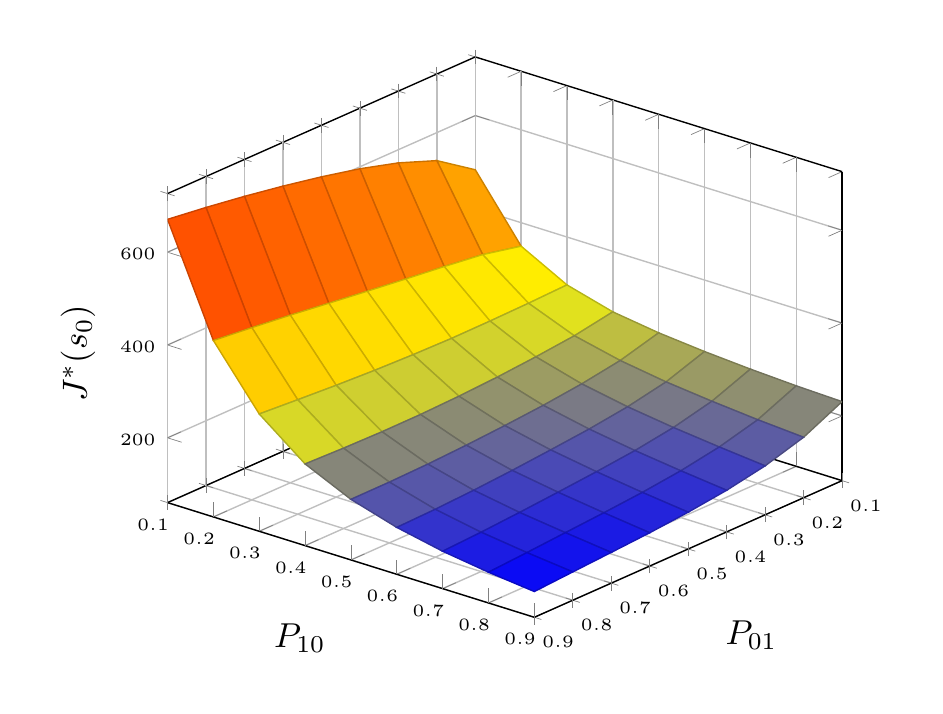}
	\caption{Impact of different combinations of stochastic process's state transition probabilities on $J^*(s_0)$.}
	\label{fig:transitionProbabilitiesVsCost}
\end{figure}

Fig.~\ref{fig:transitionProbabilitiesVsCost} presents $J^*(s_0)$ for different combinations of state transition probabilities for the stochastic process.
Probability $P_{01} (P_{10})$ in Fig.~\ref{fig:transitionProbabilitiesVsCost} is the probability for the stochastic process to make a transition from the normal (alarm) state to the alarm (normal) state at the end of a time-slot.
Probabilities $P_{00}$ and $P_{11}$ are derived by expressions $1 - P_{01}$ and $1 - P_{10}$ respectively.
The largest value for $J^*(s_0)$ occurs when $(P_{01}, P_{10})=(0.9, 0.1)$, i.e., when the stochastic process has a high probability to make a transition from the normal to the alarm state, and, once in the alarm state it has a small probability to return to the normal state.
The aforementioned cost will decrease whenever the probability to return to the normal state $P_{10}$ increases. 
One would expect a similar reduction in cost when the values of $P_{01}$ decrease, however, results in Fig.~\ref{fig:transitionProbabilitiesVsCost} reveal that decreasing $P_{01}$ may actually increase $J^*(s_0)$. 
More specifically, the lowest $J^*(s_0)$ value appears when $(P_{01}, P_{10})=(0.9, 0.9)$ and $J^*(s_0)$ actually increases as $P_{01}$ decreases from $0.9$ to $0.1$.
This may seem counter intuitive initially, since we expect that when $P_{01}$ is small, the system will spend less time in the alarm state, thus the cost $J^*(s_0)$ will be smaller. The intuition behind this phenomenon is that when both $P_{01}$ and $P_{10}$ assume large values, the stochastic process will only spend a small number of time-slots in each state. 
Given that the transmission success probability $P_s$ is large, neither $\Delta_k^0$ nor $\Delta_k^1$ will ever assume large values and thus the transition costs, as dictated by (\ref{eq:transitionCost}), will be low.

\begin{figure}[htb!]
	\centering
	\begin{subfigure}[]{0.49\columnwidth}
		\centering
		\includegraphics[width=\linewidth]{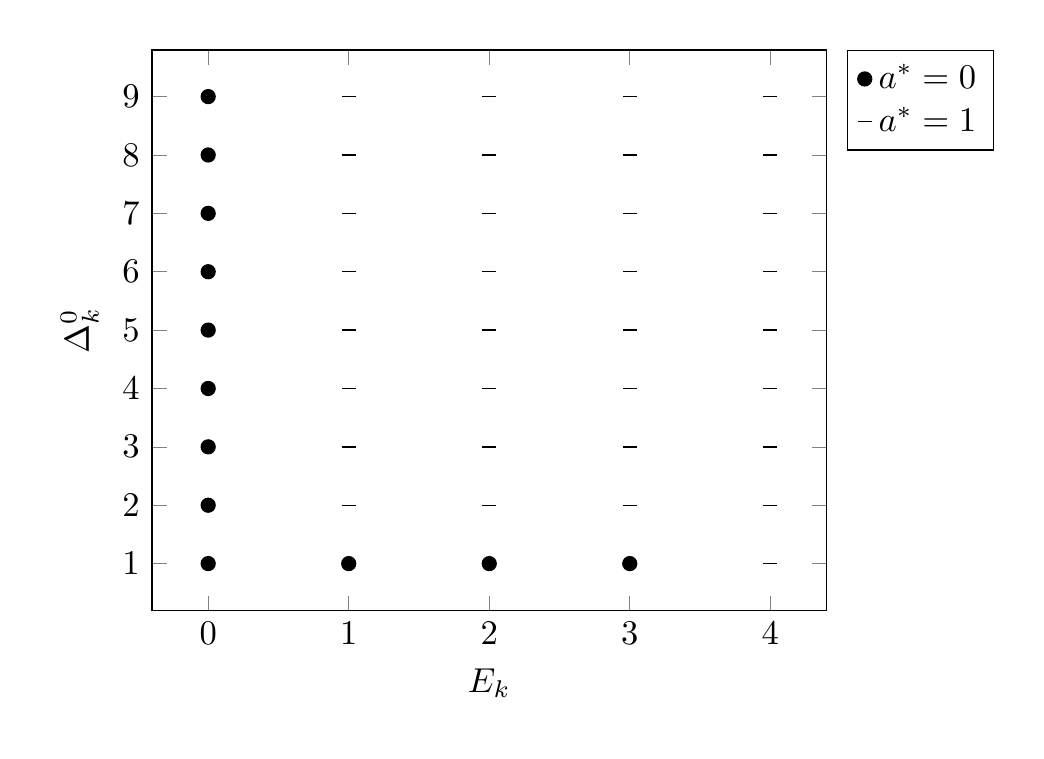}
		\caption{$Z_k = 0, P_e = 0.8$}
		\label{fig:Pe08z0}	
	\end{subfigure} 
	\begin{subfigure}[]{0.49\columnwidth}
		\centering
		\includegraphics[width=\linewidth]{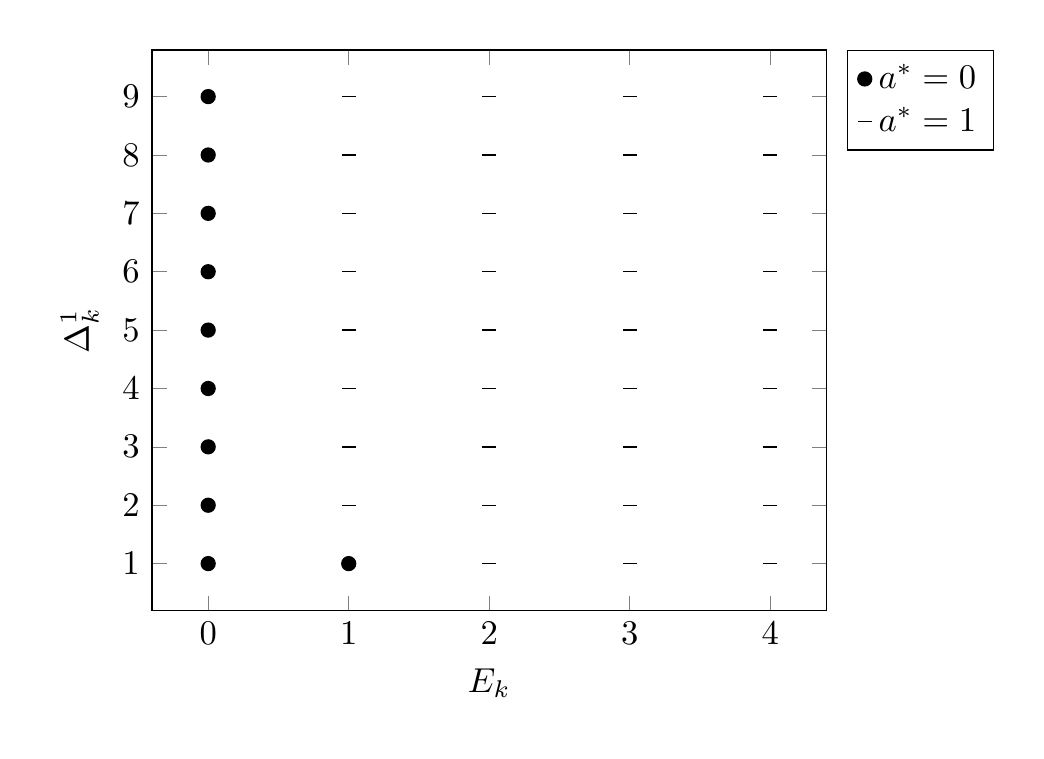}
		\caption{$Z_k = 1, P_e = 0.8$}
		\label{fig:Pe08z1}
	\end{subfigure} 
	\begin{subfigure}[]{0.49\columnwidth}
		\centering
		\includegraphics[width=\linewidth]{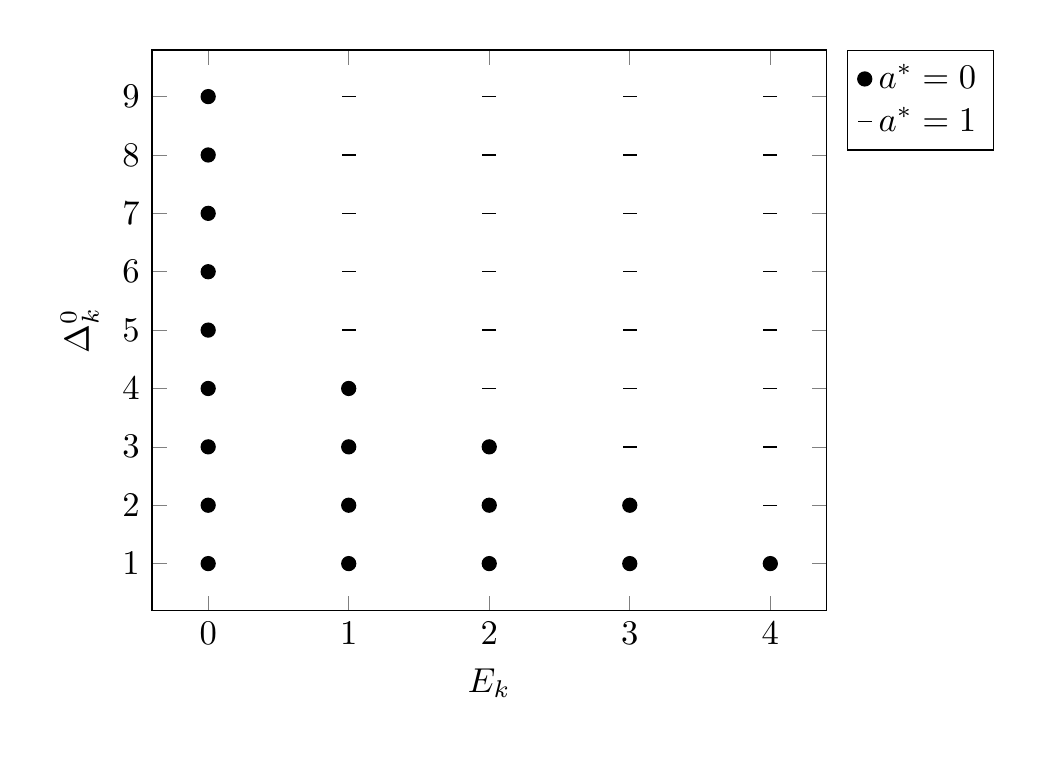}
		\caption{$Z_k = 0, P_e = 0.4$}
		\label{fig:Pe04z0}
	\end{subfigure}
	\begin{subfigure}[]{0.49\columnwidth}
		\centering
		\includegraphics[width=\linewidth]{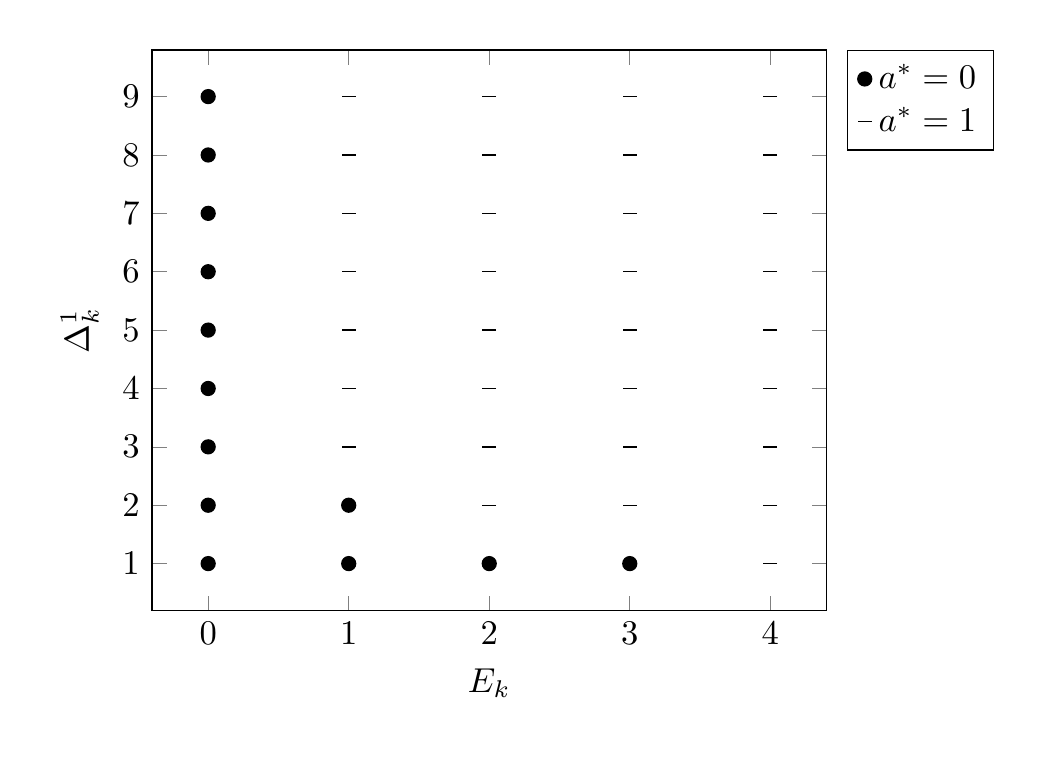}
		\caption{$Z_k = 1, P_e = 0.4$}
		\label{fig:Pe04z1}
	\end{subfigure}
	\caption{Visualization of the optimal actions when the stochastic process is in the normal state ($Z_k=0$) or the alarm state $(Z_k=1)$ and $P_e$ is either $0.8$ or $0.4$. Recall that $a^*=1$ ($a^*=0$) indicates the action of transmitting (not transmitting) a status update as dictated by optimal policy $\mu^*$.}
	\label{fig:policyActions}
\end{figure}
In Fig.~\ref{fig:policyActions} we depict the optimal policy $\mu^*$ for two scenarios.
In the first scenario energy is harvested at each time-slot with high probability, i.e., $P_e = 0.8$, while in the second one $P_e$ was set to a lower value of $0.4$.
For both experiments the transmission success probability $P_s$ was set to $0.8$, the capacity of the energy buffer was set to $5$ and the stochastic process' state transition probabilities $P_z$ were those defined in (\ref{eq:transitionProbabilityMatrix}). 
More specifically, Fig.~\ref{fig:Pe08z0} presents the optimal actions to be taken by the transmitter as a function of the number of energy units $E_k$ stored in the energy buffer and the value of the AoI counter $\Delta_k^0$ when $Z_k = 0$ and $P_e=0.8$.
Fig.~\ref{fig:Pe08z1} presents the corresponding results for the case where the stochastic process is in the alarm state, i.e. $Z_k = 1$ and
Figs.~\ref{fig:Pe04z0} and~\ref{fig:Pe04z1} present the corresponding results for the second scenario where $P_e = 0.4$.

Comparing Figs.~\ref{fig:Pe08z0} and~\ref{fig:Pe08z1} we can see that, when the probability to harvest energy is high, the actions dictated by optimal policy do not differ much between the two states of the stochastic process.
Actually, the optimal policy $\mu^*$ will still tend to reserve energy when $Z_k=0$ by not transmitting a status update ($a^*=0$) when $(E_k, \Delta_k^0) = \{(2,1), (3,1)\}$, but this is the only difference between the two cases. 
\textit{Energy reservation, in anticipation of alarm periods, becomes much more emphatic when the probability to harvest energy takes a lower value}. Comparing Figs.~\ref{fig:Pe04z0} and~\ref{fig:Pe04z1} we see that optimal policy will restrain the transmitter from sending status updates when $Z_k =0$, even when a large number of energy units is stored in the energy buffer. This is done so as to avoid the quadratic cost imposed when $Z_k^1$. The depicted optimal policy is further justified by noting the stochastic process' transition probability values.
Matrix $P_z$ indicates that once the stochastic process gets in an alarm state, it will remain in that state with high probability ($P_{11} = 0.8$). \textit{This means that reserving energy is necessary to accommodate the possibly long periods of the stochastic process being in the alarm state}.  

\section{Conclusions and Future work}
In this work, we considered an energy harvesting status update system that monitors a stochastic process which can be either in a normal or in an alarm state of operation. We formulated the problem as an MDP and derived optimal status update transmission policies for a wide range of system's configurations. We evaluated numerically the impact of the energy buffer's capacity, transmission success probability, and the stochastic process' transition probabilities on the system's efficiency. As a next step we intent to utilize Deep Reinforcement Learning techniques in order to address scenarios where the system's model is unknown and the values of $\Delta_{max}^z$ take large values, i.e., the state space becomes too large to be represented in tabular form.

\ifCLASSOPTIONcaptionsoff
\newpage
\fi

\bibliographystyle{IEEEtran}
\bibliography{bibliography}

\end{document}